# Fluid-induced pattern-formation in 3D photonic crystals with spatially varying surface functionalization

Ian B. Burgess[1*], Lidiya Mishchenko[1], Benjamin D. Hatton[1,2], Marko Lončar[1], and Joanna Aizenberg[1,2,3†]

[1]*School of Engineering and Applied Sciences, Harvard University, Cambridge MA, USA*
[2]*Wyss Institute for Biologically Inspired Engineering, Harvard University, Cambridge MA, USA*
[*]*ibburges@fas.harvard.edu*
[†]*jaiz@seas.harvard.edu*

**Abstract:** We introduce a 3D porous photonic crystal whose inner surfaces are chemically functionalized in arbitrary spatial patterns with micro-scale resolution. We use this platform to demonstrate pattern-formation.

Three dimensional (3D) photonic crystals (PCs), materials with a 3D-periodic variation in refractive index, have been the subject of extensive scientific interest since their inception over two decades ago)[1-8]. Even when there is not sufficient index-contrast to allow a complete 3D photonic bandgap, 3D PCs display exceptionally bright reflected colors arising from photonic stop gaps in particular crystal directions [8]. Structural colors from PC structures are exhibited in a wide range of biological organisms, and often display dynamic tunability [9]. Infiltration and inversion with materials that are capable of dynamic actuation has produced a broad class of PCs with structural colors that can be dynamically manipulated by mechanical force, temperature, pH, electrostatic/electrochemical forces etc. [3-8].

In this work we describe pattern formation in a large-area, optically uniform inverse-opal 3D PC [10], containing localized variations in the inner surface chemical functionality. When the opal is submerged in a fluid, the solvent penetrates the pores only in areas with compatible surface functionalization. When the pores of an inverse-opal PC are filled the refractive index contrast is diminished, redshifting and dramatically diminishing the brightness of the reflected color in these areas. For PCs with patterned inner-surface chemistry, the infiltration condition for a given fluid is satisfied in some areas, but not in others, creating a visible pattern that disappears when the PC is dried. We use this principle to inscribe multiple fluid-selective patterns into one PC that disappear when it is dried. To pattern with several functionalities, we can either add reactive groups or remove them selectively through masked introduction of reagents. Functionality patterns can also be erased and rewritten, by first being homogenized through longtime exposure to $O_2$-plasma or other etchants, followed by selective re-deposition of functional groups.

Selectively functionalized porous PCs possess the unique property of allowing selective infiltration of different fluids in different regions. Here we demonstrate this principle by using $SiO_2$ inverse opal prepared by co-assembly [10]. Selective regions of this crystal were locally silanized through a mask. Fig. 1A shows an example of selective infiltration of a PDMS prepolymer. After selective infiltration, the PDMS precursor can be polymerized inside the channels, thus allowing the visualization of wetting interfaces by scanning-electron microscopy. Fig. 1B shows an example of a PC with three rectangular regions displaying different hydrophobic surface functionalities surrounded by a background having hydrophilic surface functionality. When the entire strip is submerged in the common laboratory solvents - water, acetone and isopropyl alcohol - different optical patterns are observed, due to selective infiltration into areas with compatible surface functionalization.

Owing to its ability to induce the infiltration of solvents in a selective manner, this material also allows for spatially directed wicking. Fluid transport through the PC can occur by capillarity (wicking) or by active pumping. Fig. 1C shows and example of wicking-induced pattern formation: different liquids wick from the same starting location into different patterns. The channels defined here are treated with a hydrophilic surface functionality (vertical stripe) and a weakly hydrophobic, but alcohol-wetting functionality. The background is treated with an omniphobic functionality, providing lateral confinement for the wicking. Different patterns are revealed by liquids or liquid mixtures with varying surface properties.

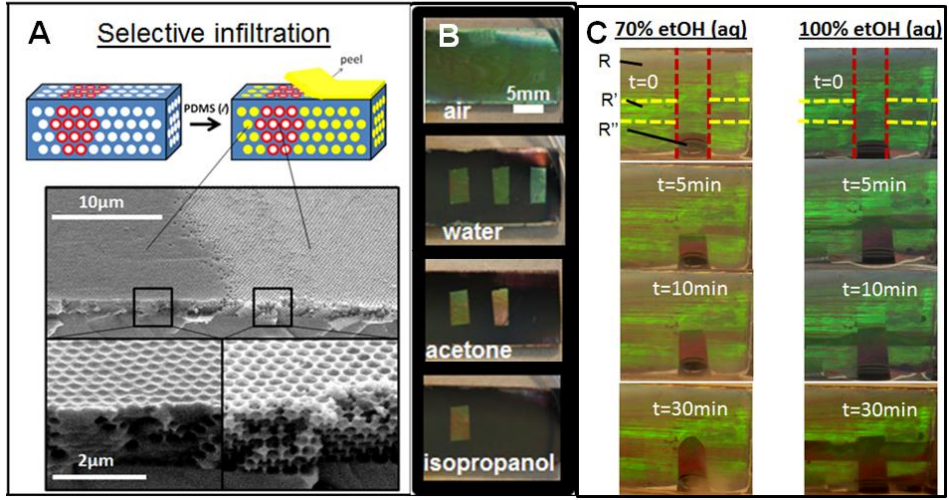

**Figure 1** - (A) Top: schematic of a 3D photonic crystal whose pores have spatially patterned surface chemistry. We visualize the patterning interfaces by selectively infiltrating a liquid that can be cured after infiltration (e.g. PDMS). Bottom: SEM images showing a typical infiltration interface. (B) A PC having a pattern of three "bars" of different hydrophobic surface chemistries on a hydrophilic background. When immersed in common laboratory solvents, different patterns appear. (C) Wicking patterns in PCs: A strongly hydrophilic channel (R'') and a slightly hydrophobic channel (R') are defined on a strongly hydrophobic background (R). Depending on the surface tension of the seed drop, the liquid wicks into different patterns. Left: 70% ethanol in water wicks only vertically. Right: pure ethanol wicks both vertically and horizontally.

In this report we have presented a photonic crystal that contains many different surface functionalities in spatially varying patterns. We have confined and directed fluids and fluid mixtures in designer patterns, all in a structurally homogeneous material. The patterned wetting is revealed by the disappearance of a bright color in the selectively infiltrated regions. We expect that this technique could also be generalized to a wide variety of porous inorganic media such as mesoporous ceramics [11,12] or silica gels, making it suitable to a host of different applications.

**Acknowledgements:**

This work was supported by Air Force Office of Scientific Research Award FA9550-09-1-0669-DOD35CAP. IBB acknowledges support from the Natural Sciences and Engineering Research Council of Canada through the PGS-D program.